\documentclass[conference]{IEEEtran}
\IEEEoverridecommandlockouts
\usepackage{cite}
\usepackage{amsmath,amssymb,amsfonts}
\usepackage{algorithmic}
\usepackage{graphicx}
\usepackage{textcomp}
\usepackage{xcolor}
\usepackage{subcaption}
\usepackage{float}
\usepackage{booktabs}
\usepackage[font=small]{caption}  
\usepackage{balance}
\usepackage{array}
\usepackage{threeparttable} 
\newcolumntype{P}[1]{>{\centering\arraybackslash}m{#1}}

\usepackage[labelformat=simple]{subcaption}

\def\BibTeX{{\rm B\kern-.05em{\sc i\kern-.025em b}\kern-.08em
    T\kern-.1667em\lower.7ex\hbox{E}\kern-.125emX}}
\begin{document}

\title{A 40 GHz Low-Power Variable-Gain Low Noise Amplifier in 28-nm CMOS Process\\
}

\author{\IEEEauthorblockN{Harshith Reddy}
\IEEEauthorblockA{\textit{Dept. of Electrical and Electronics Engineering} \\
\textit{Birla Institute of Technology and Science}\\
Pilani, India \\
f20220025@pilani.bits-pilani.ac.in}
\and
\IEEEauthorblockN{Pankaj Arora}
\IEEEauthorblockA{\textit{Dept. of Electrical and Electronics Engineering} \\
\textit{Birla Institute of Technology and Science}\\
Pilani, India \\
pankaj.arora@pilani.bits-pilani.ac.in}
}
\IEEEpubid{\makebox[\columnwidth]{%
979-8-3315-7201-3/25/\$31.00~\copyright~2025 IEEE \hfill}
\hspace{\columnsep}\makebox[\columnwidth]{}}

\maketitle

\begin{abstract}
A Low-Power Variable Gain (VG) mm-Wave Low Noise Amplifier (LNA) is designed and simulated in a 28-nm CMOS process. The LNA utilizes a simple, yet novel, technique presented in this paper to vary the small-signal output resistance to provide gain control. The amplifier also utilizes forward body biasing to reduce the supply voltage to 0.7 V and enhance power efficiency. A simultaneous noise and input matching (SNIM) technique is used to provide robust input matching and noise performance during gain adjustment. The proposed VG-LNA achieves a peak gain of 21 dB at 40.5 GHz with a noise figure of 2.8 dB and consumes only 4.5 mW. At the highest gain configuration, an input-referred 1-dB compression of -21 dBm and IP3 of -7.8 dBm are achieved, which increase to -14.8 dBm and 1.2 dBm, respectively, at the lowest gain configuration. Regardless of the gain control voltage, the LNA attains a very good FoM as compared to the state-of-the-art.
\end{abstract}

\begin{IEEEkeywords}
Low Power, mm-Wave, Variable Gain, Forward Body Biasing, Low Noise Amplifier.
\end{IEEEkeywords}

\section{Introduction}
With the recent advancements in wireless technologies, the fifth-generation (5G) systems require higher carrier frequencies that go into the mm-Wave region to support higher data transfer rates \cite{main-paper}. Concurrent fulfillment of low noise-figure (NF), low power consumption, and compact-chip area is critical \cite{Continually-Stepped}, and such demands pose major challenges for low-noise amplifiers (LNAs) as the minimum NF of a single MOS device is directly proportional to its operating frequency \cite{E-W-band-paper}. Being the first active block in a receiver chain, an LNA is considered a key component in the design of the wireless receiver frontend. A gain-control mechanism is typically needed, as for weak incoming signals, a high gain and minimum additive noise are required, and when the power level of the input signal is high, the LNA is expected to provide a low or medium gain with good linearity, such that the receiver chain won’t be saturated or distorted \cite{harvard-paper}. Conventional receiver architectures employ an LNA and a variable-gain amplifier (VGA) separately to control the overall gain of the receiver. Because of the use of two dedicated blocks, this approach increases the power consumption, area, and design complexity in the receiver chain \cite{Continually-Stepped}.
\begin{figure}
    \centering
    \includegraphics[width=0.9\linewidth]{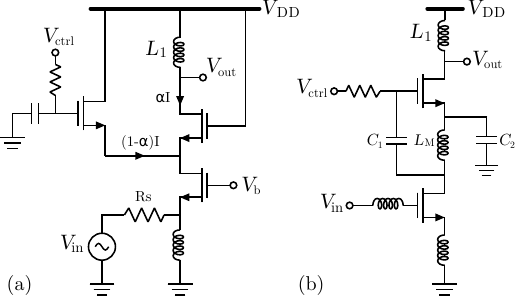}
    \caption{Examples of conventional variable-gain LNA architectures based on bias-point variation: (a) current-steering mechanism in a cascode amplifier, and (b) two-stage current-reused CS topology with gain variation through the bias voltage of the second-stage transistor.}
    \label{fig:conv-vg}
\end{figure}
There are a lot of variable-gain LNAs that have been proposed \cite{razavi} -\cite{vg-3}, as also shown in Fig. \ref{fig:conv-vg}. These methods control the bias voltages for achieving gain variation or diverting bias current to achieve the same. Such techniques could result in the biased transistors leaving the perfect operating region \cite{0.6-vg}, especially across process corners. As a consequence, gain variation could possibly hamper the input matching, noise performance, and reverse isolation. To solve such issues, a gain variation via adjusting the output resistance is proposed using an auxiliary transistor, which will be discussed in more detail subsequently.

This paper is divided into four sections. Section II analyzes in detail the working of the proposed architecture. Section III discusses the simulation results achieved. Concluding remarks are made in Section IV.

\section{Proposed Design}
\begin{figure*}[htbp]
    \centering
    \includegraphics[width=\linewidth]{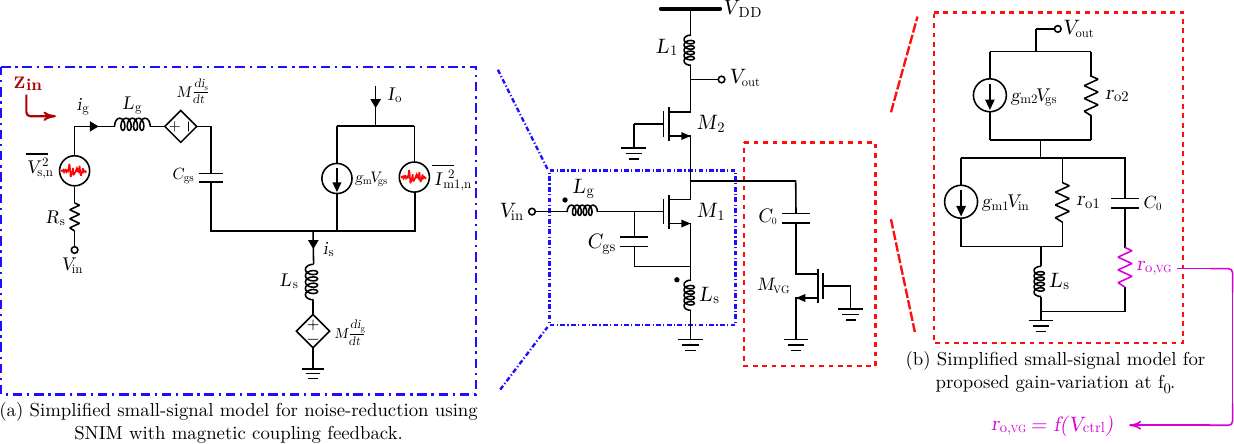}
    \caption{First-stage of proposed LNA with modifications for (a) noise-reduction and (b) variable gain.}
    \label{fig:first-stage}
\end{figure*}
The proposed LNA design is a two-stage cascode common source topology, where the first stage is modified to provide SNIM (simultaneous Noise and Input Matching) using magnetically coupled feedback proposed by Enis Kobal et al. \cite{main-paper}, and variable gain using an auxiliary transistor, as shown in Fig. \ref{fig:first-stage}. The following section has four subsections, which talk about the input matching, noise analysis, gain variability, and forward body-biasing, respectively.

\subsection{Input Impedance Matching}
Matching the input of the LNA to the source impedance is crucial for maximizing power transfer and reducing the input reflection coefficient. A badly matched input strongly degrades the noise figure because of more noise being effectively amplified to the output of the LNA instead of the required signal.

The input small-signal model seen by the source is shown in Fig. \ref{fig:first-stage}(a), with the gate connected inductor \(L_g\) and the source degeneration inductor \(L_s\) mutually coupled with a coupling coefficient \(k\) and mutual inductance of \(M=k\sqrt{L_gL_s}\). The emf generated because of mutual coupling is indicated in the small-signal model using two dependent voltage sources in series with \(L_g\) and \(L_s\), and with emf \(M\frac{di_s}{dt}=sMi_s\) and \(M\frac{di_g}{dt}=sMi_g\) respectively. The input impedance as a function of frequency can be written as
\begin{equation}
Z_{in}(s) = \frac{g_{m1}(L_s+M)}{C_{gs}} + \frac{1}{sC_{gs}} + s(L_g+L_s+2M)
    \label{eq:input-impedance}
\end{equation}
where \(g_{m1}\) is the transconductance of the transistor \(M_1\) and \(C_{gs}\) is the total gate-to-source capacitance. In order to achieve impedance matching, the real part must be 50\(\Omega\), i.e., the first term in equation \ref{eq:input-impedance}, and the imaginary part to zero, which is achieved at the resonance frequency of \(C_{gs}\) and \(L_g+L_s+2M\). The inductance of \(L_s\) can be set to achieve a 50\(\Omega\) real impedance at the frequency of resonance, which can be adjusted by tuning \(L_g\) to meet the frequency of operation.

\subsection{Noise Analysis}
Feedback techniques are often adopted in designing LNAs in order to shift the optimum noise frequency to the desired point \cite{noise-techniques}, especially in lower technology nodes where the thermal noise Power Spectral Density (PSD) increases with an increase in frequency \cite{E-W-band-paper}. The simplified small-signal model for noise analysis of the magnetic coupling feedback is shown in Fig. \ref{fig:first-stage}, where the input source noise is modelled using a noisy voltage source with PSD \(\overline{V_{s,n}^2}\) and a current source with thermal noise PSD \(\overline{I_{m1,n}^2}\) for the transistor \(M_1\). Using the theorem of superposition, the input noise source \(\overline{V_{s,n}^2}\) can be made zero to analyze the impact of feedback on the intrinsic noise generated by the circuit. So the feedback noise current flowing through the gate branch can be written as, 

\begin{equation}
i_g=\frac{-s^2C_{gs}(M+L_s)\times i_{m1,n}}{s^2C_{gs}(L_g+L_s+2M)+sg_{m1}(M+L_s)+sC_{gs}R_s+1}
    \label{eq:feedback-noise-current}
\end{equation}
Here, \(i_{m1,n}\) is the noise current of \(M_1\), notice the negative sign of the current \(i_g\) indicating negative feedback. Now, the total output noise PSD can be written as
\begin{equation}
\overline{I_{out,n}^2} 
= \overline{I_{m1,n}^2} \times
\Bigg|
\frac{
s^2C_{gs}(L_g+L_s+2M) + sC_{gs}R_s + 1}
{\scalebox{0.95}{$
\begin{aligned}
   &s^2C_{gs}(L_g+L_s+2M) \\
   &\quad +\, s\big(g_{m1}L_s + g_{m1}M + C_{gs}R_s\big) + 1
\end{aligned}$}}
\Bigg|^2
\label{eq:output-noise-current}
\end{equation}

The above two equations are a result of approximations such as ignoring the noise effects of the cascode transistor \(M_2\) and the auxiliary transistor \(M_{VG}\), which are reasonable assumptions given the high gain of \(M_1\) \cite{main-paper}. It can be observed from the equation that \(\overline{I_{out,n}^2}\) can be completely nullified at the zero frequency. Now, also adding the noise contribution of the input source, the noise factor can be derived as \cite{razavi}
\begin{equation}
F=\frac{|\overline{I_{total,n}}|^2}{|\overline{I_{s,n}}|^2}=1+\frac{\overline{I_{out,n}^2}}{\overline{I_{s,n}^2}}
    \label{eq:basic-noise-factor}
\end{equation}
\begin{equation}
F=1+\frac{\eta  \gamma (1+sC_{gs}R_s+s^2C_{gs}(L_g+L_s+2M))^2}{g_{m1}R_s}
    \label{eq:noise-factor}
\end{equation}
For,
\begin{flalign}
\overline{I_{m1,n}^2} &= 4k_BT\gamma\cdot \eta g_{m1} &&
    \label{eq:m1-noise} \\
\overline{V_{s,n}^2} &= 4k_BTR_s \text{ or, } \overline{I_{s,n}^2} = 4k_BT/R_s &&
    \label{eq:source-noise}
\end{flalign}
Here, \(k_B\), \(T\), \(R_s\) and \(\gamma\) are the Boltzmann constant, absolute temperature, source resistance and excess channel thermal noise coefficient respectively, and lastly \(\eta\) is the ratio of source transconductance \(g_{ms}\), to \(g_m\) in saturation.

The presence of coupling provides an additional degree of freedom to designers to be able to achieve higher linearity through feedback, input matching, and noise reduction. In this design, reduced parasitic effects and footprint area can be attained with the use of smaller \(L_g\) and \(L_s\) sizes, and increased frequency of operation. During design, a noise figure as low as 2.3 dB was achieved with a trade-off of reduced linearity and increased power consumption.

\subsection{Variable Gain}
\begin{figure}
    \centering
    \includegraphics[width=1\linewidth]{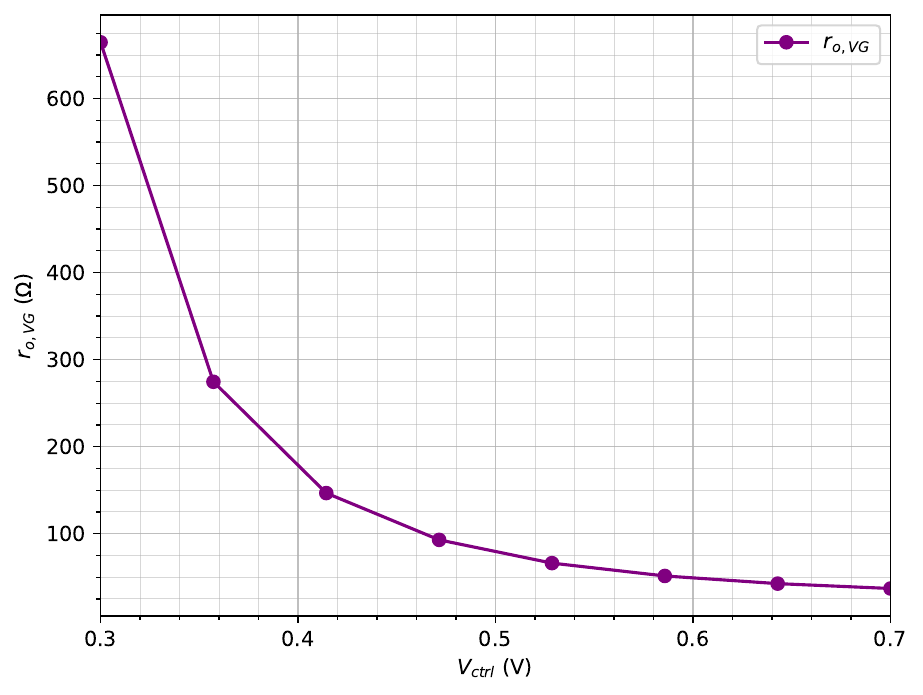}
    \caption{Simulated small-signal resistance of transistor \(M_{VG}\).}
    \label{fig:ron}
\end{figure}

As also discussed in Section I, Gain variation is utilized in LNAs to adapt to varying input signal levels, striking an effective balance between linearity and noise performance. Standard techniques control the bias voltages for achieving gain variation or diverting bias currents to achieve the same. Such techniques could result in the biased transistors leaving the perfect operating region \cite{0.6-vg}, especially across process corners. Gain variation, via such methods, could hamper input matching, noise performance, reverse isolation,  and not enhance linearity very well as the gain is reduced.
Variation of gain is proposed using an auxiliary transistor \(M_{VG}\) in series with a capacitor \(C_{0}\) as shown in Fig. \ref{fig:first-stage}. The capacitor \(C_0\) blocks DC, preserving the operating points of the whole amplifier and preventing additional power consumption. The applied control bias voltage to \(M_{VG}\) varies the small-signal resistance \(r_{o,VG}\) in an inverse relation that looks like Fig. \ref{fig:ron}. The first stage simplified small-signal model for gain calculation, ignoring magnetic feedback, is shown in Fig. \ref{fig:first-stage}(b). Approximately and intuitively, the gain at the frequency of operation can be written as
\begin{equation}
\frac{V_{out}}{V_{in}}=\frac{g_{m1}}{1+sg_{m1}L_s}\Bigg[r_{o1}||\frac{1+sC_0r_{o,VG}}{sC_0}\Bigg]\cdot\big[1+g_{m2}r_{o2}\big]
\label{eq:var-gain}
\end{equation}
where \(g_{m1}\), \(g_{m2}\), \(r_{o1}\) and \(r_{o2}\) are the small-signal transconductances and drain-to-source resistances of \(M_1\), and \(M_2\) respectively. As the control bias voltage is increased, the net output small-signal resistance decreases, thereby reducing the gain. \(C_0\) is taken large enough so that the impedance posed by it is small in comparison to \(r_{o,VG}\) and \(1/\tau\) is a decade away from the operating frequency. With \(r_{o,VG}\) going as low as 45\(\Omega\) from Fig. \ref{fig:ron}, a \(C_0\) of 0.75pF is sufficiently big enough for an operating frequency of 40 GHz. The incorporation of \(M_{VG}\) is done in the first stage because of the presence of a good noise reduction technique, instead of the second stage, because gain variation had some impact on maximum gain frequency (\(f_0\)) and was also observed to deteriorate linearity (IP3).

\begin{figure}[!b]
    \centering
    \includegraphics[width=1\linewidth]{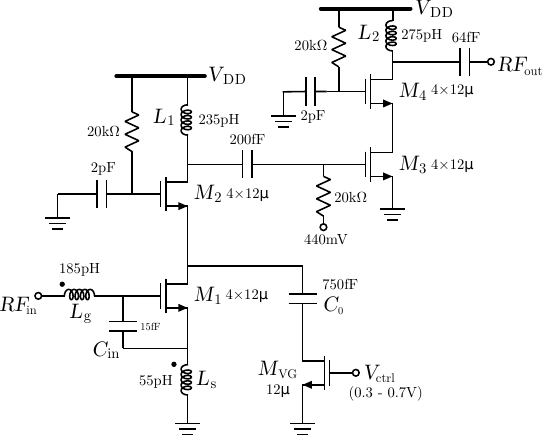}
    \caption{Proposed variable-Gain LNA schematic. (\(L_{min}\)=30nm)}
    \label{fig:schematic}
\end{figure}

\subsection{Forward Body Biasing}
Forward body biasing has been applied in the proposed design to reduce the threshold voltages of the transistors, which allows for low-\(V_{DD}\) operation and more power efficiency with higher transistor transconductances. The threshold voltage \(V_T\) for an NMOS with body effect can be expressed as follows \cite{sedra}
\begin{equation}
V_{T} = V_{T0}+\gamma\big[\sqrt{2\phi_f+V_{SB}}-\sqrt{2\phi_f}\big]
    \label{eq:body}
\end{equation}
where, \(V_{T0}\) is the threshold voltage with \(V_{SB}=0\), \(\gamma\) is the body effect coefficient and \(\phi_f\) is the bulk Fermi potential. Reducing  \(V_T\) allows cascode transistors with a low-\(V_{DD}\) supply to operate with stronger inversion, thereby improving noise performance, as compared to weakly inverted transistors, which suffer higher gate-induced noise \cite{fbb}. In the proposed design, with a body bias voltage of 0.55 V, the threshold voltages of the transistors were brought down from 0.46 V to 0.41 V. The bias voltage was kept below 0.7 V to avoid forward biasing the substrate p-n junction, restricting body leakage current to 1.2\(\mu\)A, and ensuring safe operation. Apart from a slight trade-off on IP3 and P1dB, and maintaining almost similar performance characteristics, the power consumption was reduced from 5.9 mW with a 0.9 V supply to 4.5 mW with a 0.7 V supply.

\begin{figure}[t]
    \centering
    \includegraphics[width=1\linewidth]{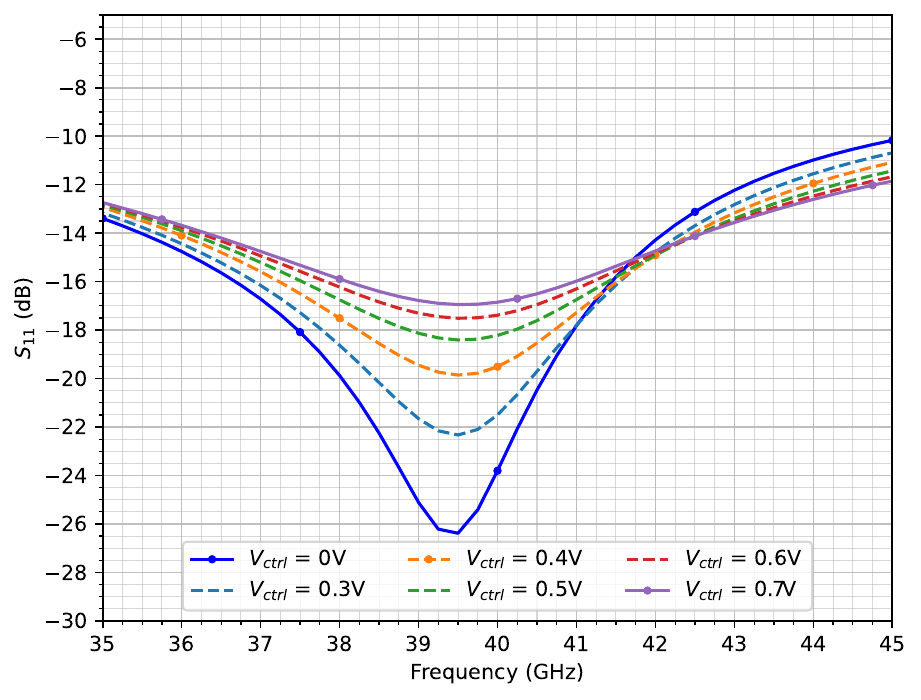}
    \caption{Input Match of the proposed LNA, with varying gain control.}
    \label{fig:s-params}
\end{figure}

\begin{figure}[t]
    \centering
    \includegraphics[width=1\linewidth]{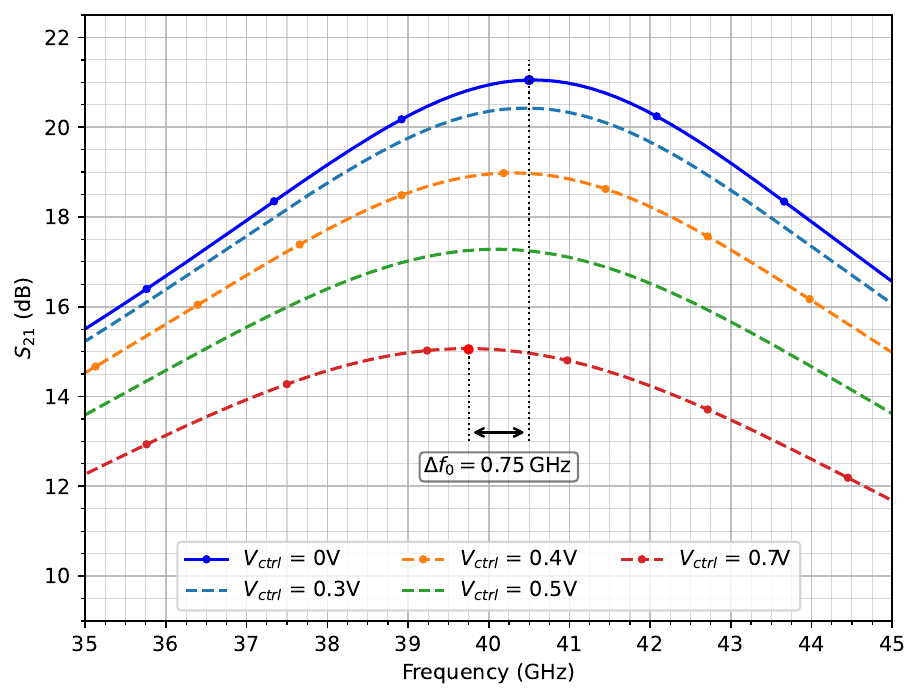}
    \caption{Gain variation (\(S_{21}\)) vs. Frequency for different \(V_{ctrl}\).}
    \label{fig:gain}
\end{figure}
\begin{figure}
    \centering
    \includegraphics[width=1\linewidth]{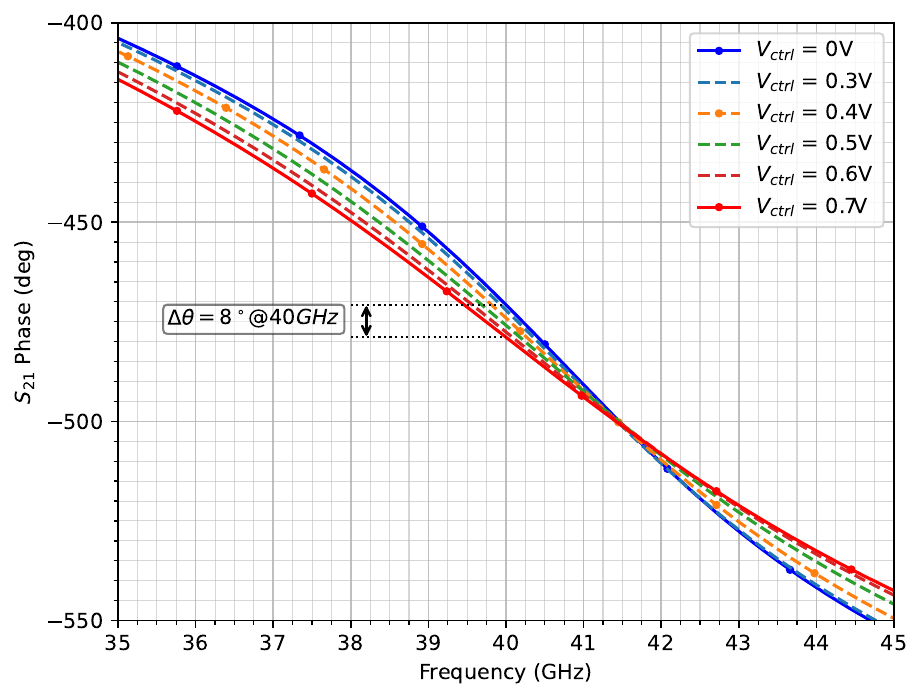}
    \caption{\(S_{21}\) Phase vs. Frequency variation for different \(V_{ctrl}\).}
    \label{fig:phase}
\end{figure}
\begin{figure}[t]
    \centering
    \includegraphics[width=1\linewidth]{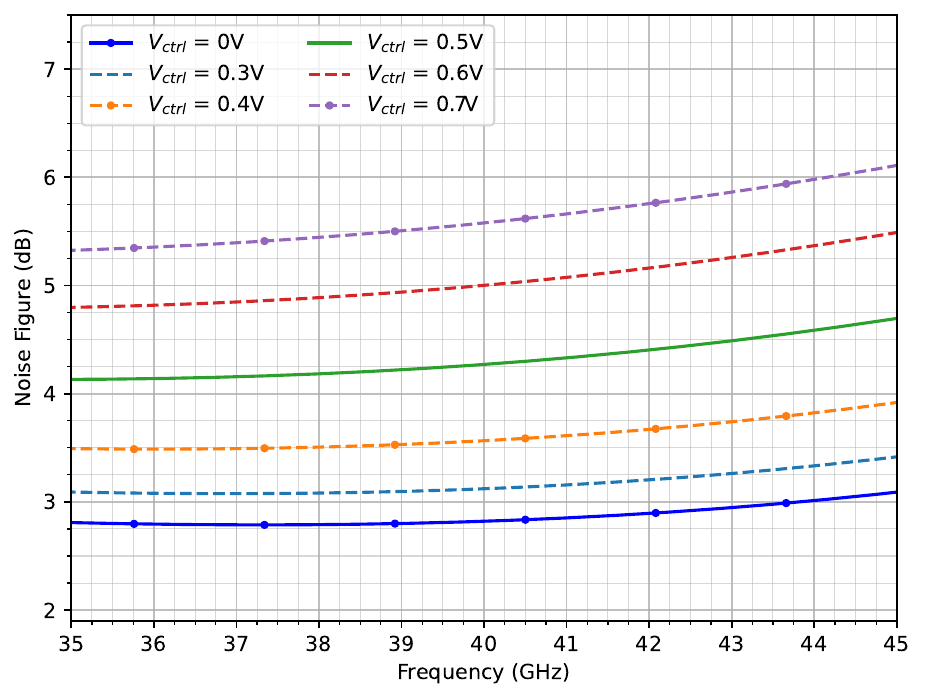}
    \caption{Noise-Figure of LNA vs. Frequency for different \(V_{ctrl}\)}
    \label{fig:NF}
\end{figure}

\section{Results Obtained}
The proposed variable-gain LNA was designed in UMC 28-nm CMOS technology, with the circuit schematic and sizings of each component shown in Fig. \ref{fig:schematic}. The LNA was designed and simulated in Cadence Virtuoso. It operates using a 0.7 V supply and consumes very little power of, only 4.5 mW, with forward biasing of 0.55 V. Fig. \ref{fig:s-params} shows the simulated \(S_{11}\) vs. frequency for the proposed LNA. It has a minimum \(S_{11}\) of -26.3 dB at 40 GHz and good matching in the range of 34 GHz to 45 GHz of under -10 dB. It is well matched with the input even during gain transitions, and this gain variation can be incorporated in single stage LNAs, unlike many similar ones altering gain in the second or third stage.

Fig. \ref{fig:gain} shows the gain (\(S_{21}\)) variation with the control bias voltage \(V_{ctrl}\). The LNA achieves a maximum gain of 21 dB at 40.5 GHz with zero \(V_{ctrl}\) and a minimum gain of 15 dB at 39.75 GHz with 0.7 V \(V_{ctrl}\), presenting only a maximum difference of 0.75 GHz in the maximum gain frequency. At the highest gain, the LNA has a 3-dB bandwidth between 37 GHz to 43.8 GHz, which is a span of 6.8 GHz, and at the lowest gain, the LNA has a 3-dB bandwidth between 34.8 GHz and 44.6 GHz, which is a span of 9.8 GHz.

Fig. \ref{fig:phase} shows the \(S_{21}\) phase variation across different \(V_{ctrl}\) biases. The LNA presents a maximum of 8\textdegree deviation at 40 GHz, and less than one degree at 41.5 GHz. Fig. \ref{fig:NF} shows the simulated noise figure plots for each \(V_{ctrl}\) applied. At the highest gain, the LNA presents a very good noise figure of 2.8 dB, and a noise figure of 5.5 dB at the lowest gain. The simulated input-referred 1-dB compression point (\(P_{1dB}\)) at 42 GHz with the highest gain is -21 dBm and goes up to -14.8 dBm at the lowest gain bias.

\begin{table*}[!t]
\centering
\scriptsize
\begin{threeparttable}
\caption{Performance Comparison of the proposed LNA with the State-of-the-Art}
\label{tab:comparison}
\begin{tabular}{|P{1.85cm}|P{0.8cm}|P{1.5cm}|P{1.1cm}|P{0.6cm}|P{0.6cm}|P{0.5cm}|P{1.7cm}|P{1.7cm}|P{0.45cm}|P{0.6cm}|P{1cm}|}
\hline
\textbf{Reference} & 
\textbf{Process} & 
\textbf{Topology} & 
\textbf{3-dB BW (GHz)} & 
\textbf{\(\mathbf{f_0}\) (GHz)} & 
\textbf{Max-Gain (dB)} & 
\textbf{NF (dB)} & 
\textbf{IIP3 (dBm)} & 
\textbf{P-1dB (dBm)} & 
\textbf{VDD (V)} & 
\textbf{Power (mW)} &
\textbf{FoM (dB)} \\ 
\hline
\cite{main-paper} TCAS-II’23 & 22-nm FDSOI & 2-stage cascode & 23.7–28.5 & 25.7 & 23.1 & 2.1 & –16.5 & –17.7  & 0.6 & 5.6 & 24.22 \\
\hline
\cite{E-W-band-paper} TMTT’20 & 22-nm FDSOI & 3-stage cascode & 71–83\tnote{*} & 77 & 20 & 4.6 & –17.8\tnote{**} & –27.4  & 1 & 9 & 39.1 \\
\hline
\cite{ims-25-paper} IMS’25 & 40-nm & 3-stage differential CS & 50–70.9 & 62 & 23.1 & 3.8 & –4.9** & –14.5 & 1.1 & 33 & 65.37 \\
\hline
\cite{tmtt-24} TMTT’24 & 65-nm & CS + differential cascode & 28.1–48.1 & 38.5\tnote{*} & 21.5 & 2.7 & –7.6 & Not reported & 1 & 22 & 60 \\
\hline
\cite{jssc-25} JSSC’25 & 110-nm & 2-stage current reuse CS & 6.4–7.8 & 7\tnote{*} & 13.6 & 2.8 & –5.2 & Not reported & 1 & 1.2 & 35.99 \\
\hline
\cite{rfic-21} RFIC’21 & 28-nm & 2-stage cascode + diff CS & 22.2–43 & 30\tnote{*} & 21.1 & 3.5 & –3 & –16 & 0.9 & 22.3 & 63.26 \\
\hline
\cite{tmtt-20-caltech} TMTT’20 & 22-nm FDSOI & 2-stage cascode & 22.5–27.6 & 24.8\tnote{*} & 23.2 & 2.38 & –10.4 & –21 & 0.8 & 5.5 & 55.56 \\
\hline
\cite{tcas-24} TCAS-II’24 & 40-nm & 2-stage current reuse cascode + diff CS & 50.6–67 & 60 & 16.8 & 4.4 & -3.4\tnote{**} & -13 & 1.2 & 33 & 51.43 \\
\hline
\textbf{This Work (highest gain)} & \textbf{28-nm} & \textbf{2-stage cascode} & \textbf{37–43.8} & \textbf{40.5} & \textbf{21} & \textbf{2.8} & \textbf{–7.8} & \textbf{–21} & \textbf{0.7} & \textbf{4.5} & \textbf{63.15} \\
\hline
\textbf{This Work (lowest gain)} & \textbf{28-nm} & \textbf{2-stage cascode} & \textbf{34.8–44.6} & \textbf{39.75} & \textbf{15} & \textbf{5.5} & \textbf{1.2} & \textbf{–14.8} & \textbf{0.7} & \textbf{4.5} & \textbf{63.02} \\
\hline
\end{tabular}

\begin{tablenotes}
\scriptsize
\item[*] Estimated from plot. 
\item[**] Approximated using IIP3 = P1dB + 9.6 dB.
\end{tablenotes}

\end{threeparttable}
\end{table*}

\begin{figure}
    \centering
    \includegraphics[width=1\linewidth]{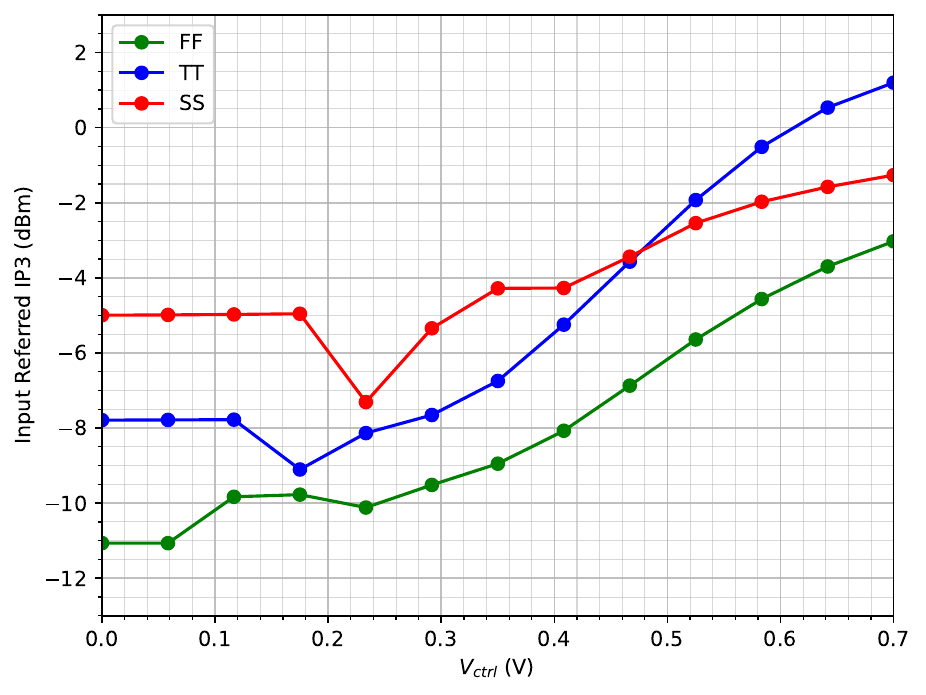}
    \caption{Input-Referred IP3 vs. \(V_{ctrl}\) across process corners at 42 GHz.}
    \label{fig:ip3}
\end{figure}

Fig. \ref{fig:ip3} shows the variation of simulated input-referred third-order intercept point (IIP3) vs. control voltage \(V_{ctrl}\) across different process corners for the frequency of 42 GHz. At the highest gain, IIP3 is -7.8 dBm, and 1.2 dBm at the lowest gain state.   It can be also seen that the IIP3 is not heavily affected by process variation. At the FF corner, increases from -11 dBm to -3 dBm, and -5 dBm to -1.2 dBm at the SS corner, respectively, as the control voltage \(V_{ctrl}\) is increased. It can be inferred that the proposed LNA does not have a strong linearity deviation with process variations, but has a strong correlation with gain reduction and control.

\begin{figure}[t]
    \centering
    \includegraphics[width=1\linewidth]{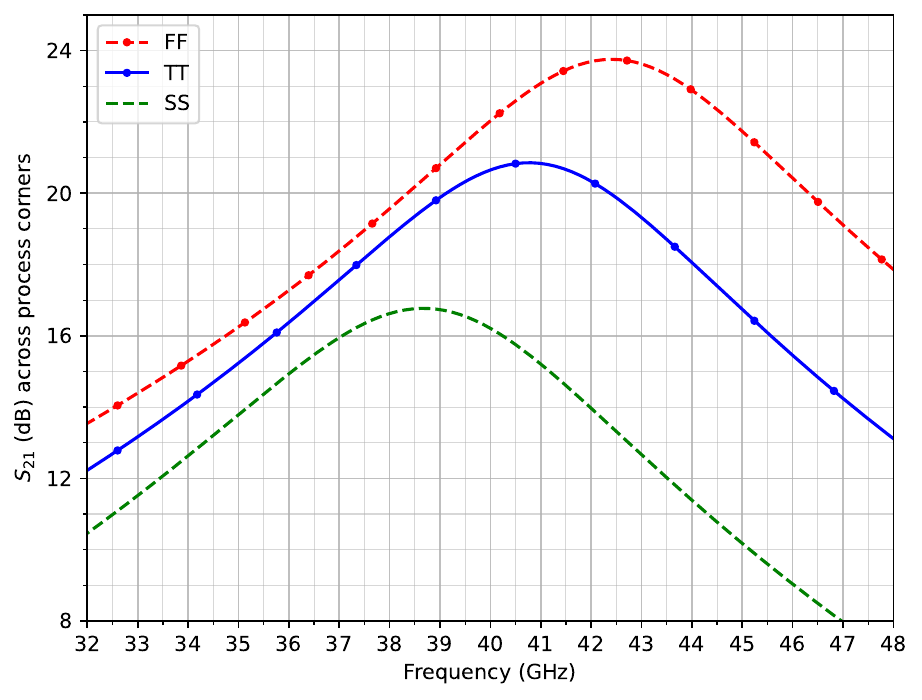}
    \caption{\(S_{21}\) of LNA vs. Frequency across process corners, \(V_{ctrl} = 0\).}
    \label{fig:gain-corners}
\end{figure}

\begin{figure}
    \centering
    \includegraphics[width=1\linewidth]{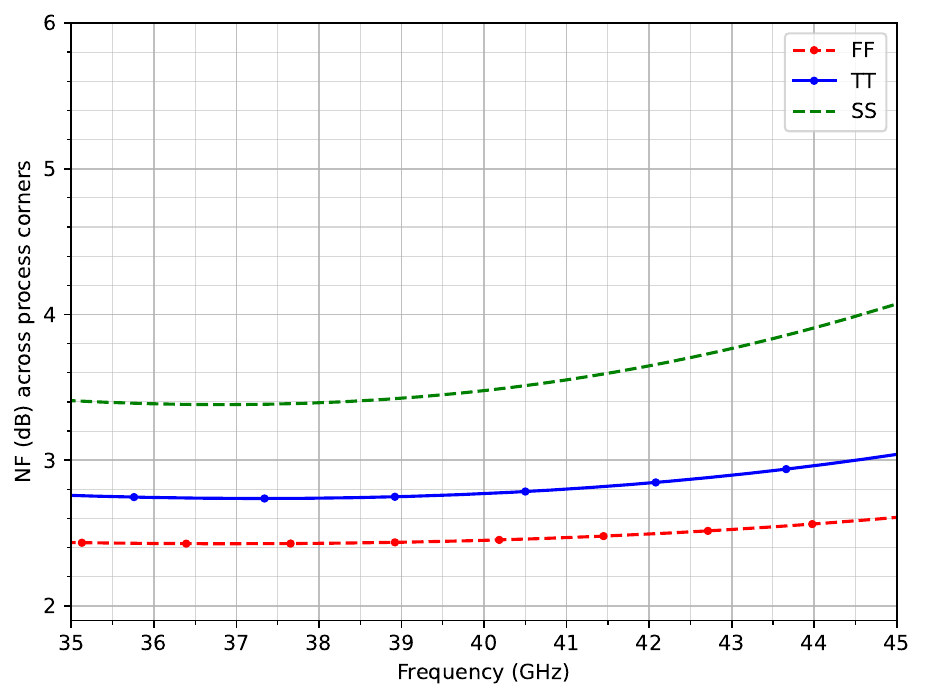}
    \caption{Noise-Figure of LNA across process corners, \(V_{ctrl} = 0\).}
    \label{fig:nf-corners}
\end{figure}

Fig. \ref{fig:gain-corners} and Fig. \ref{fig:nf-corners} show the variation of gain (\(S_{21}\)), and noise figure across process corners when \(V_{ctrl}\) is zero, i.e., at maximum gain. At the slowest corner (SS), the maximum gain frequency drops to 38.8 GHz with a maximum gain of 16.8 dB and a noise figure of 3.6 dB. At the fastest corner (FF), the maximum gain frequency increases to 42.3 GHz with a maximum gain of 23.8 dB and a noise figure of 2.45 dB. At the SS corner, the power consumption of the LNA reduces to 3.2 mW and increases to 5.9 mW at the FF corner.

Table \ref{tab:comparison} compares the proposed LNA with the state-of-the-art, with the expression calculated for the Figure of Merit (FoM) as
\[FoM=20log\bigg[\frac{Gain(Abs.)\cdot BW_{3dB}\cdot f_0\cdot IIP3(mW)}{(F-1)\cdot P_{DC}(mW)}\bigg]\]

At the highest gain, the proposed LNA has an FoM of 63.15 dB and 63.02 dB at the lowest gain, which suggests that the LNA has very little variation in overall FoM and performance with gain adjustment. A few recent works, like \cite{ims-25-paper} and \cite{rfic-21} from Table \ref{tab:comparison}, have comparable FoM to this work, because they are not variable-gain LNAs and have at the minimum three stages as compared to only a 2-stage cascode of this work, hence exceeding the expected footprint area of this work. The proposed LNA is anticipated to have a lower footprint area because of its compact 2-stage design with small inductors and transistor sizes as compared to all the recent works. In addition, this work has a better noise figure and IP3 linearity as compared to all works, while having a very low power consumption.

\section{CONCLUSION}
This paper presents a narrowband low-power variable-gain mm-Wave LNA, designed and simulated in the UMC 28-nm CMOS process. It utilizes a Simultaneous Noise and Input Matching (SNIM) technique using two gate-to-source magnetically coupled feedback inductors. It also uses forward body biasing to reduce the threshold voltages of the transistors, allowing for a lower supply voltage and reduced power consumption. The LNA's gain control is proposed by adjusting the small-signal output resistance of the first stage, using an auxiliary transistor connected to ground and in series with a capacitor that blocks DC and prevents hampering of the amplifier's bias points. This method is power-efficient, and unlike conventional methods, it is more robust against process corners, and it prevents transistors from leaving their perfect operating region, which could inhibit linearity enhancement while gain is being reduced. The simulation results of the proposed VG-LNA have been presented, with the performance summary and comparison with recent works. The VG-LNA achieves a very good FoM as compared to the state-of-the-art, without compromise, as gain is reduced. As discussed earlier in this section, the significant enhancement in IIP3 can be seen from reducing gain, from -7.8 dBm (0.166 mW) at the highest gain to 1.2 dBm (1.32 mW) at the lowest gain. It has also been shown by simulations that the IIP3 performance is still very good across process corners. The LNA achieves a high frequency of operation, with a 3dB bandwidth of 37-43.8 GHz, and a peak gain of 21 dB with a noise figure of 2.8 dB at the highest gain. The gain can be reduced all the way to 15.04 dB using a control voltage between 0.3 and 0.7 V, and producing minimal phase variations. The proposed LNA is very power efficient and area efficient,  being only 2-stage with reduced inductor and transistor sizes, and consumes only 4.5 mW with a supply of 0.7 V.

Future work will focus on the layout implementation of the proposed design to validate the performance through post-layout simulations and the reduction in footprint area.

\section*{Acknowledgment}
The authors would like to thank the Chips to Startup (C2S) program, Ministry of Electronics and Information Technology (MeitY), Government of India, for providing access to the necessary EDA tools and resources that enabled this work.

\vspace{12pt}

\end{document}